\documentclass[a4paper,aps,twocolumn,nofootinbib]{revtex4}
\RequirePackage[colorlinks,hyperindex]{hyperref}
\RequirePackage[english]{babel}
\RequirePackage[latin1]{inputenc}
\RequirePackage[T1]{fontenc}
\RequirePackage{mathrsfs}
\RequirePackage{amsmath}
\RequirePackage{amssymb}
\RequirePackage{amsbsy}
\RequirePackage{color}
\RequirePackage{bm}
\RequirePackage{graphicx}
\RequirePackage{slashed}
\hypersetup{colorlinks=true,breaklinks=true,urlcolor=blue,linkcolor=red}
\begin{document}
%%%%%%%%%%%%%%%%%%%%%%%%%%%%%%%%%%%%%%%%%%%%%%%%%%%%%%%%%%%%%%%%%%%%%%%%%%%%%%%%%%%%%%%%%%%%%%%%%%%
\title{\bf{Cross section for Bhabha and Compton scattering beyond quantum field theory}}
\author{Flora Moulin\footnote{moulin@lpsc.in2p3.fr}$^{\hbar}$, 
Luca Fabbri\footnote{fabbri@dime.unige.it}$^{\nabla}$, 
Aur\'{e}lien Barrau\footnote{barrau@lpsc.in2p3.fr}$^{\hbar}$}
\affiliation{$^{\hbar}$Laboratoire de Physique Subatomique et de Cosmologie, 
Universit\'{e} Grenoble-Alpes CNRS/IN2P3,\\
53 avenue des Martyrs, 38026 Grenoble cedex, FRANCE \\
$^{\nabla}$DIME Sezione di Metodi e Modelli Matematici, Universit\`{a} degli studi di Genova,\\
via all'Opera Pia 15, 16145 Genova, ITALY}
\date{\today}
%%%%%%%%%%%%%%%%%%%%%%%%%%%%%%%%%%%%%%%%%%%%%%%%%%%%%%%%%%%%%%%%%%%%%%%%%%%%%%%%%%%%%%%%%%%%%%%%%%%
\begin{abstract}
We consider the theory of spinor fields written in polar form, that is the form in which the spinor components are given in terms of a module times a complex unitary phase respecting Lorentz covariance. In this formalism, spinors can be treated in their most general mathematical form, without the need to restrict them to plane waves. As a consequence, calculations of scattering amplitudes can be performed by employing the most general fermion propagator, and not only the free propagator usually employed in QFT. In this article, we use these quantities to perform calculations in two notable processes, the electron-positron and Compton scatterings. We show that although the methodology differs from the one used in QFT, the final results in the two examples turn out to give no correction as predicted by QFT.
\end{abstract}
%%%%%%%%%%%%%%%%%%%%%%%%%%%%%%%%%%%%%%%%%%%%%%%%%%%%%%%%%%%%%%%%%%%%%%%%%%%%%%%%%%%%%%%%%%%%%%%%%%%
\maketitle
%%%%%%%%%%%%%%%%%%%%%%%%%%%%%%%%%%%%%%%%%%%%%%%%%%%%%%%%%%%%%%%%%%%%%%%%%%%%%%%%%%%%%%%%%%%%%%%%%%%
\section{Introduction}
Quantum field theory (QFT) is a magnificent theory in its predictive power. It has successfully passed all experimental tests put forward up to now. Its core consists in taking plane-wave solutions of the free equations, promoting fields to be operators and expanding scattering amplitudes in terms of radiative corrections that are supposed to encode the interactions.

In this framework, predictions for the anomaly of the magnetic moment of leptons and for the hyper-fine splitting of hydrogen-like atoms have been confirmed to an astonishing precision. The philosophy that lies at the foundations of QFT is that, in the perturbative expansion, all propagators are given for the free fields, the full information about the interactions being encoded by loops of virtual particles. Naively, one may see the process as a Taylor expansion of the given interaction.

Nevertheless, the mathematical structure still needs a proper definition \cite{Peskin}: its inconsistencies range from the fact that equal-time commutation relationships may lack a precise definition \cite{streater} to the fact that the interaction picture used for the perturbative expansion may not (and in fact in some cases does not) exist \cite{haag}. Even worse may be the fact that all calculations are done in terms of plane-wave solutions, which are not square-integrable, so that strictly speaking they cannot be physically acceptable.

In this respect, one may wonder if, to describe a given interaction, it could be possible to get rid of the perturbative series of loop corrections and free propagators, and try to describe the interaction in a whole new way.

More precisely, the question we intend to ask is: given that QFT is built on free propagators and vertices encoding the interaction, what would happen if we were to neglect vertices but employ propagators that already contain the interaction? That is, is it possible to consider only the tree-level case but with interacting propagator and still get the results of QFT?

To investigate this question one should use a formalism that can encompass interactions in the propagator, and this can only be done by taking into account the most general propagator that is possible.

In order to get the general propagator for spinors, we need to make a considerable use of the so called polar form, the form in which spinor fields can be written in such a way that each of their components is a module times a complex unitary phase. Generally speaking, this process spoils manifest covariance. However, recent developments made it possible to use an explicitly covariant approach. We thus take advantage of this formalism, and the ensuing polar form, for the calculation of the spinor propagator together with the scattering amplitude.

After that, we will apply such an expression to see whether QFT results can be recovered or not.

The processes in which we are interested are the Bhabha (electron-positron) scattering and the Compton scattering in the specific case where the electron is in a hydrogen-like potential. 
%%%%%%%%%%%%%%%%%%%%%%%%%%%%%%%%%%%%%%%%%%%%%%%%%%%%%%%%%%%%%%%%%%%%%%%%%%%%%%%%%%%%%%%%%%%%%%%%%%%
\section{General Spinor Theory}
To treat spinors, we begin by introducing the Clifford matrices $\boldsymbol{\gamma}^{\mu}$ satisfying
\begin{eqnarray}
&\{\boldsymbol{\gamma}^{a}, \boldsymbol{\gamma}^{b} \} \!=\! 2\eta^{ab}\mathbb{I}
\end{eqnarray}
known as Clifford algebra. In terms of this
\begin{eqnarray}
& \boldsymbol{\sigma}^{ab} \!=\! \frac{1}{4} 
\left[\boldsymbol{\gamma}^{a}\!,\!\boldsymbol{\gamma}^{b}\right]
\end{eqnarray}
are the generators of the complex Lorentz group
\begin{eqnarray}
&\boldsymbol{S}\!=\!e^{\frac{1}{2}\theta_{ab}\boldsymbol{\sigma}^{ab}}
\end{eqnarray}
where $\theta_{ab}=-\theta_{ba}$ are the parameters of the real Lorentz transformation $\Lambda$ mentioned above. The relation
\begin{eqnarray}
&2i\boldsymbol{\sigma}_{ij}\!=\!\varepsilon_{ijpq}\boldsymbol{\pi}\boldsymbol{\sigma}^{pq}
\end{eqnarray}
defines the $\boldsymbol{\pi}$ matrix\footnote{This matrix is what is usually referred to as $\boldsymbol{\gamma}_{5}$ but because in the $4$-dimensional space-time this index has no meaning we prefer to use a notation in which no index appears at all.}\!\!, whose existence ensures by means of Schur's lemma that the transformation $\boldsymbol{S}$ is reducible.

A spinor field $\psi$ is defined as a vector field in the space of spin, or complex Lorentz transformations. It is a ``column'' of $4$ complex scalars satisfying
\begin{eqnarray}
&\psi\!\rightarrow\!\boldsymbol{S}\psi
\end{eqnarray}
as a general transformation law. Its adjoint $\overline{\psi}\!=\!\psi^{\dagger}\boldsymbol{\gamma}_{0}$ is defined in such a way that the spinorial bi-linears
\begin{eqnarray}
&2i\overline{\psi}\boldsymbol{\sigma}^{ab}\psi\!=\!M^{ab},\\
&\overline{\psi}\boldsymbol{\gamma}^{a}\boldsymbol{\pi}\psi\!=\!S^{a},\\
&\overline{\psi}\boldsymbol{\gamma}^{a}\psi\!=\!G^{a},\\
&i\overline{\psi}\boldsymbol{\pi}\psi\!=\!\Theta,\\
&\overline{\psi}\psi\!=\!\Phi,
\end{eqnarray}
are all real tensorial quantities.

The dynamical character is determined by the Dirac spinor field equation
\begin{eqnarray}
&i\boldsymbol{\gamma}^{\mu}\boldsymbol{\nabla}_{\mu}\psi\!-\!m\psi\!=\!0,
\end{eqnarray}
where $m$ is the mass of the field.

Using the bi-linear quantities, one may perform the Lounesto classification of spinor fields \cite{L}. In this classification, spinors are divided into six major classes \cite{Cavalcanti:2014wia}, which are split as in the following: if $\Theta$ and $\Phi$ do not both vanish identically the spinors are called \emph{regular} and they constitute the classes I, II, III (the actual distinction between these three classes is physically irrelevant for us in the present paper). Instead, if $\Theta\!\equiv\!\Phi\!\equiv\!0$ the spinors are called \emph{singular} and split in further sub-classes according to the following: if $S^{a}$ and $M^{ab}$ do not both vanish identically the spinors are called \emph{flag-dipole} and they constitute class IV; if $S^{a}\!=\!0$ the spinors are called \emph{flagpole} and they constitute class V; if $M^{ab}\!=\!0$ the spinors are called \emph{dipole} and they constitute class VI. Classes I, II, III altogether contain the Dirac spinors, and while no physical example of class IV is known we still have that class V contains Majorana spinors and class VI contains Weyl spinors. Consequently, it is not generally possible to dismiss these classes of singular spinors \cite{daSilva:2012wp}. However, in the present article our interest will be focused on regular spinors. For these it was shown \cite{Fabbri:2016msm} that it is always possible to write them in chiral representation as
\begin{eqnarray}
&\!\psi\!=\!\phi e^{-\frac{i}{2}\beta\boldsymbol{\pi}}
\boldsymbol{S}\left(\!\begin{tabular}{c}
$1$\\
$0$\\
$1$\\
$0$
\end{tabular}\!\right)
\label{spinor}
\end{eqnarray}
where $\phi$ is called the module and $\beta$ is the Yvon-Takabayashi angle, and in which $\boldsymbol{S}$ is a generic spin transformation. Via a direct calculation we can see that
\begin{eqnarray}
&\Theta\!=\!2\phi^{2}\sin{\beta}\label{b2}\\
&\Phi\!=\!2\phi^{2}\cos{\beta}\label{b1}
\end{eqnarray}
showing that $\phi$ and $\beta$ are one scalar field and one pseudo-scalar field. Because in general we have the validity of the Fierz identity $G^{a}G_{a}\!=\!-S_{a}S^{a}\!=\!\Theta^{2} \!+\! \Phi^{2}$ then it is also possible to have the vectors $G^{a}$ and $S^{a}$ normalized into the vectors $g^{a}$ and $s^{a}$ according to
\begin{eqnarray}
&S^{a}\!=\!2\phi^{2}s^{a}\\
&G^{a}\!=\!2\phi^{2}g^{a}
\end{eqnarray}
as it is possible to check. It is also possible to see that
\begin{eqnarray}
&M_{ab}\!=\!2\phi^{2}(g^{j}s^{k}\varepsilon_{jkab}\cos{\beta}\!+\!g_{[a}s_{b]}\sin{\beta})
\end{eqnarray}
showing that only the vector and axial-vector with scalar and pseudo-scalar are independent and therefore that the $\phi$ and $\beta$ are the only true real degrees of freedom of the spinorial field. Because we can always boost into the frame in which $G^{a}$ loses its spatial part we conclude that $G^{a}$ must be the velocity of the spinor. A straightforward approximation in non-relativistic limit shows that $S^{a}$ is the spin of the spinor. So $\boldsymbol{S}$ is the transformation that takes the most general spinor into the rest frame where the spin is aligned along the third axis. We remark that (\ref{spinor}) is unique up to the reversal of the third axis, encoding the difference between spin-up and spin-down spinor eigen-states. It is however possible to get the alternative
\begin{eqnarray}
&\!\psi\!=\!\phi e^{-\frac{i}{2}\beta\boldsymbol{\pi}}
\boldsymbol{S}\left(\!\begin{tabular}{c}
$-1$\\
$0$\\
$1$\\
$0$
\end{tabular}\!\right)
\end{eqnarray}
which simply corresponds to the transformation $\psi\!\rightarrow\!\boldsymbol{\pi}\psi$ encoding the passage from the spinor describing a particle to the corresponding spinor describing its antiparticle \cite{Peskin}.

The spinorial field consists in $8$ real components, and the polar form (\ref{spinor}) makes their meaning clear \cite{Fabbri:2016msm}: the scalar $\phi$ is what in non-relativistic quantum mechanics gives the amplitude of probability while $\beta$ describes the dynamics between right-handed and left-handed chiral parts thus vanishing in non-relativistic limit. The three space components of velocity and spin are described as rapidities and angles in terms of the parameters of the $\boldsymbol{S}$ transformation and thus can always be transferred to the underlying space-time structure, as we will see later.

Among other useful relationships, one should mention
\begin{eqnarray}
&\!\!\!\!\psi\overline{\psi}\!\equiv\!\frac{1}{2}
\phi^{2}[(g_{a}\boldsymbol{\mathbb{I}}\!+\!s_{a}\boldsymbol{\pi})\boldsymbol{\gamma}^{a}
\!\!+\!e^{-i\beta\boldsymbol{\pi}}(\boldsymbol{\mathbb{I}}
\!-\!2g_{a}s_{b}\boldsymbol{\sigma}^{ab}\boldsymbol{\pi})]
\label{F}
\end{eqnarray}
which is valid in the most general case, and in terms of which we can see that the spin-sum relationships are
\begin{eqnarray}
&\sum_{\mathrm{spin}}\psi\overline{\psi}\!\equiv\!
\phi^{2}(g_{a}\boldsymbol{\gamma}^{a}\!+\!e^{-i\beta\boldsymbol{\pi}}),
\label{spinsum}
\end{eqnarray}
where the sum is performed on all spin states \cite{Fabbri:2017pwp}.

By considering the polar form (\ref{spinor}) and since in general
\begin{eqnarray}
&\boldsymbol{S}\partial_{\mu}\boldsymbol{S}^{-1}\!=\!i\partial_{\mu}\theta\mathbb{I}
\!+\!\frac{1}{2}\partial_{\mu}\theta_{ij}\boldsymbol{\sigma}^{ij},\label{Lorentz}
\end{eqnarray}
where $\theta$ is a generic complex phase and $\theta_{ij}\!=\!-\theta_{ji}$ are the six parameters of the Lorentz group, we can define
\begin{eqnarray}
&\partial_{\mu}\theta_{ij}\!-\!\Omega_{ij\mu}\!\equiv\!R_{ij\mu}\label{R}\\
&\partial_{\mu}\theta\!-\!qA_{\mu}\!\equiv\!P_{\mu}\label{P}
\end{eqnarray}
with the gauge potential $qA_{\mu}$ and the spin connection $\Omega_{ij\mu}$. Because equations (\ref{R}, \ref{P}) contain the same information than the gauge potential and the spin connection but are proven to be real tensors, they are called the gauge-invariant vector momentum and the tensorial connection. Writing the spinor field in polar form thus consists in re-arranging the components so as to isolate the real degrees of freedom from the components that can be transferred through the frame into the underlying space-time structure where they combine with the gauge potential and the connection leading to (\ref{P}, \ref{R}). Using those variables, the spin covariant derivative is given by
\begin{eqnarray}
&\boldsymbol{\nabla}_{\mu}\psi\!=\!(\nabla_{\mu}\ln{\phi}\mathbb{I}
\!-\!\frac{i}{2}\nabla_{\mu}\beta\boldsymbol{\pi}
\!-\!iP_{\mu}\mathbb{I}\!-\!\frac{1}{2}R_{ij\mu}\boldsymbol{\sigma}^{ij})\psi
\label{decspinder}
\end{eqnarray}
in the most general case. We therefore have 
\begin{eqnarray}
&\nabla_{\mu}s_{i}\!=\!R_{ji\mu}s^{j}\label{ds}\\
&\nabla_{\mu}g_{i}\!=\!R_{ji\mu}g^{j}\label{dg}
\end{eqnarray}
as general identities on the velocity and spin.

Dirac spinor field equations are equivalently given as
\begin{eqnarray}
&B_{\mu}\!-\!2P^{\iota}g_{[\iota}s_{\mu]}
\!+\!\nabla_{\mu}\beta\!+\!2s_{\mu}m\cos{\beta}\!=\!0,\label{dep1}\\
&R_{\mu}\!-\!2P^{\rho}g^{\nu}s^{\alpha}\varepsilon_{\mu\rho\nu\alpha}
\!+\!2s_{\mu}m\sin{\beta}\!+\!\nabla_{\mu}\ln{\phi^{2}}\!=\!0,\label{dep2}
\end{eqnarray}
in which $\frac{1}{2}\varepsilon_{\mu\alpha\nu\iota}R^{\alpha\nu\iota}\!=\!B_{\mu}$ and $R_{\mu a}^{\phantom{\mu a}a}\!=\!R_{\mu}$ for the sake of simplicity. It may be interesting to highlight that from the Dirac equations one may obtain the Gordon decomposition $\nabla_{\nu}G^{\nu}\!=\!0$ which is just the expression of the conserved vector current due to the proportionality between velocity and current given by $J_{\nu}\!=\!qG_{\nu}$ as known in electrodynamic. Similarly we have that $\nabla_{\nu}S^{\nu}\!=\!2m\Theta$ as partially-conserved axial-vector current due to the proportionality between spin and axial-vector current.

The Dirac spinorial field equations are $8$ real equations that, in polar form, are converted into $2$ vector equations that specify all the space-time derivatives of the two real degrees of freedom. Notice that the angle $\beta$ is the phase-shift between the chiral parts and, as such, it encodes the information about the mass term \cite{Fabbri:2016laz}.

One can also to prove that
\begin{eqnarray}
&\!\!\!\!\!\!\!\!R^{i}_{\phantom{i}j\mu\nu}\!=\!-(\nabla_{\mu}R^{i}_{\phantom{i}j\nu}
\!-\!\!\nabla_{\nu}R^{i}_{\phantom{i}j\mu}
\!\!+\!R^{i}_{\phantom{i}k\mu}R^{k}_{\phantom{k}j\nu}
\!-\!R^{i}_{\phantom{i}k\nu}R^{k}_{\phantom{k}j\mu})\label{Riemann}\\
\!\!\!\!&qF_{\mu\nu}\!=\!-(\nabla_{\mu}P_{\nu}\!-\!\nabla_{\nu}P_{\mu})\label{Maxwell}
\end{eqnarray}
leading to the Maxwell strength and Riemann curvature of the underlying gauge and space-time structures. A gauge-covariant type of electrodynamic information is therefore encoded in (\ref{P}) while (\ref{R}) encodes a covariant type of gravitational and inertial acceleration. From (\ref{Riemann},\ref{Maxwell}), one can see that the physical information is still the one that enters the strength and curvature, so that the non-zero solutions of equations
\begin{eqnarray}
&\nabla_{\mu}R^{i}_{\phantom{i}j\nu}
\!-\!\!\nabla_{\nu}R^{i}_{\phantom{i}j\mu}
\!\!+\!R^{i}_{\phantom{i}k\mu}R^{k}_{\phantom{k}j\nu}
\!-\!R^{i}_{\phantom{i}k\nu}R^{k}_{\phantom{k}j\mu}=0\\
&\nabla_{\mu}P_{\nu}\!-\!\nabla_{\nu}P_{\mu}=0
\end{eqnarray}
describe a covariant type of gauge potential and inertial acceleration that are not related to sources \cite{Fabbri:2018crr}.

For electrodynamics such a strengthless gauge potential is related to the Aharonov-Bohm effect. A similar phenomenon is expected in the gravitational sector for the curvatureless covariant inertial acceleration \cite{Fabbri:2019kyd}.

By combining the Dirac equations in polar form (\ref{dep1},\ref{dep2}) it is possible to establish a link between $P_{a}$ and $R_{ijk}$ as
\begin{eqnarray}
&P^{\mu}\!=\!m\cos{\beta}g^{\mu}\!-\!y_{k}s^{[k}g^{\mu]}\!-\!x_{k}s_{j}g_{i}\varepsilon^{kji\mu},
\label{momentum}
\end{eqnarray}
having set $x_{k}\!=\!\frac{1}{2}(\nabla_{k}\ln{\phi^{2}}+R_{k})$ and 
$y_{k}\!=\!\frac{1}{2}(\nabla_{k}\beta+B_{k})$ for the sake of simplifying the form of the expression \cite{Fabbri:2019tad}.

To establish the most complete form of the spinor propagator, we write the Dirac equation in polar form as
\begin{eqnarray}
&(F_{k}\boldsymbol{\gamma}^{k}\!+\!y_{k}\boldsymbol{\gamma}^{k}\boldsymbol{\pi}\!-\!m)\psi\!=\!0,
\end{eqnarray}
where $F_{k}\!=\!P_{k}+ix_{k}$ in terms of the momentum. The propagator is hence the solution of the equation
\begin{eqnarray}
&(F_{k}\boldsymbol{\gamma}^{k}\!+\!y_{k}\boldsymbol{\gamma}^{k}\boldsymbol{\pi}\!-\!m)G
\!=\!\mathbb{I},
\end{eqnarray}
and is found to be given by
\begin{eqnarray}
\nonumber
&G\!=\!(|F^{2}\!+\!y^{2}\!+\!m^{2}|^{2}-|2F\!\cdot\!y|^{2}-4F^{2}m^{2})^{-1}\\
\nonumber
&\times[F_{k}\boldsymbol{\gamma}^{k}-y_{k}\boldsymbol{\gamma}^{k}\boldsymbol{\pi}-m]\\
&\times[(F^{2}+y^{2}+m^{2})\mathbb{I}+2F\!\cdot\!y\boldsymbol{\pi}+2mF_{a}\boldsymbol{\gamma}^{a}],
\label{G}
\end{eqnarray}
as proven by a direct substitution \cite{Fabbri:2020nmx}.
%%%%%%%%%%%%%%%%%%%%%%%%%%%%%%%%%%%%%%%%%%%%%%%%%%%%%%%%%%%%%%%%%%%%%%%%%%%%%%%%%%%%%%%%%%%%%%%%%%%
\section{Reduction to QFT}
As mentioned in the introduction, the computations are carried out in QFT by considering plane-wave solutions of the fundamental equations. The implementation of this requirement is performed by using
\begin{eqnarray}
&i\boldsymbol{\nabla}_{\mu}\psi\!=\!P_{\mu}\psi,
\end{eqnarray}
which has to be compared with the general form (\ref{decspinder}). It is easy to see that
\begin{eqnarray}
&(\nabla_{\mu}\ln{\phi}\mathbb{I}
\!-\!\frac{i}{2}\nabla_{\mu}\beta\boldsymbol{\pi}
\!-\!\frac{1}{2}R_{ij\mu}\boldsymbol{\sigma}^{ij})\psi=0
\end{eqnarray}
which has to be true for any spinor field. Because $\boldsymbol{\sigma}^{ij}$, $\mathbb{I}$ and $\boldsymbol{\pi}$ are linearly independent, we must have $R_{ij\mu}=0$ with $\phi$ and $\beta$ constant. As a constant pseudo-scalar has to vanish, we get that QFT requires $\nabla_{\nu}\phi$ with $\beta$ and $R_{ij\mu}$ all equal to zero. When the spin is averaged out, picking the usual normalization condition $\phi^{2}\!=\!m$ gives 
\begin{eqnarray}
&\psi\overline{\psi}\!\equiv\!\frac{1}{2}(P_{a}\boldsymbol{\gamma}^{a}\!+\!m\mathbb{I})
\end{eqnarray}
or 
\begin{eqnarray}
&\sum_{\mathrm{spin}}\psi\overline{\psi}\!\equiv\!P_{a}\boldsymbol{\gamma}^{a}\!+\!m\mathbb{I}
\end{eqnarray}
as the completeness relationships we would have encounter in the standard treatment of QFT.

The Dirac equations (\ref{dep1}, \ref{dep2}) reduce to the simpler
\begin{eqnarray}
&P^{\iota}g_{[\iota}s_{\mu]}\!-\!s_{\mu}m\!=\!0\\
&P^{\rho}g^{\nu}s^{\alpha}\varepsilon_{\mu\rho\nu\alpha}\!=\!0
\end{eqnarray}
which can be worked out to be equivalent to 
\begin{eqnarray}
&P^{\mu}\!=\!mg^{\mu}
\label{momentumsmall}
\end{eqnarray}
in general. However, a set of four conditions can not imply the validity of a system of eight equations unless some information is lost. In this case, the lost information is the one involving the internal dynamics. In fact, if it were possible that $P^{\mu}\!=\!mg^{\mu}$ then any boost in the rest frame, which is always possible for a massive particle, would also mean a boost into the frame in which the non-relativistic limit is recovered exactly. But boosting into rest frame does not mean boosting into the frame in which the non-relativistic approximation is accurate because even in their rest frame particles still have an internal dynamics. Thus it is only when the spin is neglected that the proportionality between velocity and momentum can hold, as we have in QFT.

For the propagator we would have
\begin{eqnarray}
&G\!\approx\!(P^{2}\!-\!m^{2})^{-1}(P_{k}\boldsymbol{\gamma}^{k}+m\mathbb{I})
\end{eqnarray}
singular on-shell. Now we can give a partial interpretation of QFT singular behaviour at the poles, and that is the singularity comes from neglecting the spin, or more in general the internal structure. This is to be expected for point-like particles such as those used in QFT.

Notice that the vanishing of the Yvon-Takabayashi angle and of the tensorial connection might appear as reasonable assumptions. Nevertheless, their validity leads to disturbing consequences, such as the fact that the Dirac-Maxwell field equations have no solution \cite{Fabbri:2018dje}.

That tensorial connection and Yvon-Takabayashi angle are quite generally different from zero is clear from the fact that they are non-zero in some remarkable situations such as those given by the two integrable potentials, the hydrogen atom and the harmonic oscillator \citep{Fabbri:2019kyd}.

Consequently, it would be wise to leave these quantities different from zero, working with a form of spin-sum, momentum and propagator more general than in QFT.
%%%%%%%%%%%%%%%%%%%%%%%%%%%%%%%%%%%%%%%%%%%%%%%%%%%%%%%%%%%%%%%%%%%%%%%%%%%%%%%%%%%%%%%%%%%%%%%%%%%
\section{Scattering Processes}
We now proceed to calculate the transition amplitude for the Bhabha and Compton scatterings. 

\subsection{Bhabha scattering}
To begin with, we consider an electron-positron scattering, which is simpler since, in this case, the propagator is a photon and no correction is taken.

For this process, we focus on the two main usual Feynman diagrams. The matrix element is the sum 
\begin{eqnarray}
i \mathcal{M}_{fi}= i \mathcal{M}_{fi}^s + i \mathcal{M}_{fi}^t,
\label{Mfi}
\end{eqnarray}
where $s$ and $t$ refer to the photon channels. We will compute the two channels separately, although calculations are analogous. The momenta are given by (\ref{momentum}).

The calculation of the amplitude leads to
\begin{eqnarray}
\mathcal{M}_{fi}^s &=&\bigg[ \overline{u(q_1)}(i e \gamma ^\nu) v(q_2)\bigg] \bigg[ \frac{\eta_{\mu \nu}}{s} \bigg] \bigg[\overline{v(p_2)}(i e \gamma ^\mu) u(p_1)\bigg] \nonumber \\
&=& - \frac{e^2}{s} \overline{u(q_1)} \gamma _{\mu} v(q_2) \overline{v(p_2)} \gamma ^{\mu} u(p_1),
\end{eqnarray}
and to
\begin{eqnarray}
&\overline{\mathcal{M}_{fi}^s} =- \frac{e^2}{s} \overline{u(p_1)} \gamma ^{\nu} v(p_2) \overline{v(q_2)} \gamma _{\nu} u(q_1),
\end{eqnarray}
as it is well known in QFT. However, differently from what is done in standard QFT, we consider a more general spin-sum: for the particle spinor $u$ we have
\begin{eqnarray}
\nonumber
&\sum_{\mathrm{spin}} \!u \overline{u}\!=\!\phi^{2}(g_{a}\boldsymbol{\gamma}^{a}
\!+\!e^{-i\beta\boldsymbol{\pi}})\\
&\approx\phi^{2}[\slashed{g}\!+\!\mathbb{I}(1\!-\!\beta^{2}/2)
\!-\!i\beta \boldsymbol{\pi}]
\label{approxuubar}
\end{eqnarray}
while as already mentioned the antiparticle spinor $v$ is defined by $v=\boldsymbol{\pi}u$ and hence
\begin{eqnarray}
\nonumber
&\sum_{\mathrm{spin}} \!v \overline{v}\!=\!\phi^{2}(g_{a}\boldsymbol{\gamma}^{a}
\!-\!e^{-i\beta\boldsymbol{\pi}})\\
&\approx\phi^{2}[\slashed{g}\!-\!\mathbb{I}(1\!-\!\beta^{2}/2)
\!+\!i\beta \boldsymbol{\pi}],
\label{approxvvbar}
\end{eqnarray}
up to second order in $\beta$, which is the level of approximation we consider in this section of the paper.

From these spin-sums, we get the transition amplitudes
\begin{eqnarray}
\nonumber
&\frac{1}{4} \sum _{\mathrm{spin}}|\mathcal{M}^s|^{2}=\frac{e^4\phi^8}{s^2}\mathrm{Tr}\bigg[[\slashed{h_1} + \mathbb{I}(1\!-\!\beta^{2}/2) - i \beta \boldsymbol{\pi}] \gamma _\mu \\
\nonumber
&[\slashed{h_2} - \mathbb{I} (1\!-\!\beta^{2}/2) + i \beta \boldsymbol{\pi}]\gamma _\nu \bigg]\\
\nonumber
&\times\mathrm{Tr}\bigg[[ \slashed{g_2} - \mathbb{I}(1\!-\!\beta^{2}/2) + i \beta \boldsymbol{\pi}] \gamma ^\mu \\
&[ \slashed{g_1} + \mathbb{I}(1\!-\!\beta^{2}/2) - i \beta \boldsymbol{\pi}]\gamma ^\nu \bigg],
\end{eqnarray}
and
\begin{eqnarray}
\nonumber
&\frac{1}{4} \sum _{\mathrm{spin}}|\mathcal{M}^t|^{2}=\frac{e^4 \phi ^8}{t^2}\mathrm{Tr}\bigg[[ \slashed{h_1} + \mathbb{I} (1\!-\!\beta^{2}/2) - i \beta \boldsymbol{\pi}] \gamma _\mu \\
\nonumber
&[ \slashed{g_1} + \mathbb{I} (1\!-\!\beta^{2}/2) - i \beta \boldsymbol{\pi}]\gamma _\nu \bigg] \\
\nonumber
&\times\mathrm{Tr}\bigg[[ \slashed{g_2} - \mathbb{I}(1\!-\!\beta^{2}/2) + i \beta \boldsymbol{\pi}] \gamma ^\mu \\
&[\slashed{h_2} - \mathbb{I} (1\!-\!\beta^{2}/2)+ i \beta \boldsymbol{\pi}]\gamma ^\nu \bigg],
\end{eqnarray}
with $g$ and $h$ the velocities associated with the momenta $p$ and $q$ respectively. It is easy to check that QFT results are recovered if we normalize $\phi^{2}\!=\!m$ with $p=mg$ and $q=mh$ with $\beta\!=\!0$. Calculating the traces shows that, up to a second-order correction, all $\beta$ terms disappear and
\begin{eqnarray}
\frac{1}{4} \sum _{\mathrm{spin}} | \mathcal{M}^s| ^2 = \frac{8 e^4 \phi ^8}{s^2} \bigg[ (g_2 \!\cdot\! h_1) (g_1 \!\cdot\! h_2) \nonumber \\ + (g_2 \!\cdot\! h_2)(g_1 \!\cdot\! h_1) + (g_1 \!\cdot\! g_2) + (h_1\!\cdot\! h_2) + 2\bigg]
\end{eqnarray}
with
\begin{eqnarray}
\frac{1}{4} \sum _{\mathrm{spin}} | \mathcal{M}^t| ^2 = \frac{8 e^4 \phi ^8}{t^2} \bigg[ (g_2 \!\cdot\! h_1) (g_1 \!\cdot\! h_2) \nonumber \\ + (g_2 \!\cdot\! h_2)(g_1 \!\cdot\! h_1) - (h_1 \!\cdot\! g_1) - (h_2 \!\cdot\!g_2)+ 2\bigg],
\end{eqnarray}
showing no difference with respect to the result of standard QFT. We have also explicitly checked that this remains true for the interference term. 

\subsection{Compton scattering}
We now move to the Compton scattering. This process is more complicated because it involves the evaluation of traces that contain the propagator of the fermion which, in our case, are given by intricate expressions. The generalization appears through corrections in the Yvon-Takabayashi angle. We consider an incident photon with momentum $k^a$ and an electron in a hydrogen potential with momentum $p^a$ given by (\ref{momentum}). The final particles are a photon with $k^{'a}$ and an electron with $p^{'a}$.

We still consider the two usual main Feynman diagrams. The matrix element in the $s$-channel is given by
\begin{eqnarray}
&\!\!\!\!\!\!\!\!\frac{1}{4}\sum _{\mathrm{spin}}\!|\mathcal{M}^{s}|^{2}
\!=\!\frac{e^{4}}{4}\mathrm{Tr}\!\left[S(p')\gamma^{\nu}G(q)\gamma^{\mu}
S(p)\gamma_{\mu}\overline{G(q)}\gamma_{\nu}\right]
\end{eqnarray}
as it is well known with $ q = p+k $.

For the spin-sums we take first-order corrections
\begin{eqnarray}
S(p)=\phi^2\bigg(\slashed{g}+\mathbb{I}-i\beta\boldsymbol{\pi} \bigg)
\end{eqnarray}
and
\begin{eqnarray}
S(p')=\phi^{'2}\bigg(\slashed{g}'+\mathbb{I}-i\beta'\boldsymbol{\pi}\bigg)=m\slashed{g}'
+m \mathbb{I}
\end{eqnarray}
since $\beta'$ is the Yvon-Takabayashi angle of the scattered electron, which can be considered as free. For $\beta$, one has to consider the Yvon-Takabayashi angle of the electron orbiting the nucleus, which in spherical coordinates is
\begin{equation}
\beta=-\arctan{\left( \frac{\alpha}{\Gamma}\cos{\theta}\right)},
\label{betah}
\end{equation}
with $\alpha$ the fine-structure constant and $\Gamma\!=\!\sqrt{1\!-\!\alpha^2}$ for hydrogen-like atoms. The values of the modules are given by $\phi^{'2}\!=\!m$ and $\phi^{2}=mr^{2(\Gamma-1)}e^{-2\alpha mr}/\Delta$ in which we have $\Delta(\theta)\!=\!(1\!-\!|\alpha\sin{\theta}|^{2})^{-\frac{1}{2}}$ a function of the elevation angle, as it has been found in reference \cite{Fabbri:2018crr}. It is quite an interesting fact that $\beta$, $\ln{\phi}$ and their derivatives are of the same order than the fine-structure constant.

The propagators are given by (\ref{G}) and the first-order truncation leads to
\begin{eqnarray}
\nonumber
&G(q)\!\approx\!(q^{2}\!-\!m^{2}+4q\!\cdot\!xi)^{-1}\\
&\times\bigg[W\mathbb{I} + X_a \boldsymbol{\gamma}^a + Y_a \boldsymbol{\gamma}^a \boldsymbol{\pi} + Z_{ab} \boldsymbol{\sigma}^{ab} \boldsymbol{\pi} \bigg],
\end{eqnarray}
with 
\begin{eqnarray}
W=m + \frac{2 i m (x \cdot q) }{q^2-m^2},\\
X_a =q_a + ix_a + \frac{2 i (x \cdot q)}{q^2-m^2 } q_a,\\
Y_a =\frac{1}{q^2-m^2 }[-(q^2+m^2) y_a + 2 (y \cdot q) q_a],\\
Z_{ab} =\frac{4m}{q^2-m^2 } y_a q_b.
\end{eqnarray}
It is worth noticing that starting from our general expressions and taking the first-order perturbative expansion, we obtain the imaginary term $4q\!\cdot\!x i$ in the denominator of the scalar factor without the necessity to postulate it as done in QFT. This term is needed to remove the poles and it is naturally present in the most general formalism that we are using in the present article.

At first-order in $x$, $y$, and $\beta$ we have 
\begin{eqnarray}
&W(q)W^*(q) =m^2,\\
&X(q)X^*(q) =q^2,\\
&W(q)X^*_a(q)+W^*(q)X_a(q)= 2mq_a,\\
&W(q)X^*_a(q)-W^*(q)X_a(q)= - 2\mathrm{i} m x_a,
\end{eqnarray}
so that 
\begin{eqnarray}
\nonumber
\frac{1}{4} \sum _{\mathrm{spin}} | \mathcal{M}^s| ^2 =
\frac{4 \phi ^2 m e^4}{(q^2-m^2)^2}\\
\nonumber
\times \bigg[4m^2 + 4 q^2 
-4m(g\!\cdot\!q) - q^2 (g \!\cdot\! g')\\
-4m(g'\!\cdot\! q) + m^2 (g \!\cdot\! g') + 2 (g \!\cdot\! q) (g'\!\cdot\! q) \bigg] \ \
\end{eqnarray}
for the $s$-channel. For the $u$-channel
\begin{eqnarray}
\frac{1}{4} \sum _{\mathrm{spin}} | \mathcal{M}^u| ^2 = 
\frac{4 \phi ^2 m e^4}{(f^2-m^2)^2} \nonumber \\
\times \bigg[4m^2 + 4 f^2 
-4 m (g \!\cdot\! f) - f^2 (g \!\cdot\! g') \nonumber \\ - 4 m (g' \!\cdot\! f) + m^2 (g \!\cdot\! g') + 2 (g \!\cdot\! f) (g' \!\cdot\! f) \bigg], \ \
\end{eqnarray}
as it is easy to check, with $f= p-k'$. For the interference term, we obtain
\begin{eqnarray}
\nonumber
\frac{1}{4} \sum _{\mathrm{spin}} \Re( \mathcal{M}^s \overline{\mathcal{M}^u} ) = \frac{4 \phi ^2 m e^4}{(f^2-m^2)}
\frac{1}{(q^2-m^2)}\\
\nonumber
\times\bigg[ -2m^2+ m^2(g \cdot g') +\\
+ m (q+ f ) \cdot(g+ g') +
 (q \cdot f) (1- 2(g \cdot g') ) \bigg].
\end{eqnarray} 
We should notice that the physical changes are not due to the new propagator but to the Coulomb potential, which is naturally accounted for in this framework.

\begin{figure}
\includegraphics[width=.8\linewidth]{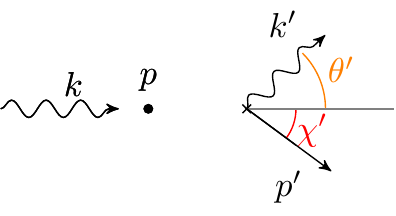} 
\caption{Compton scattering in the lab frame.}
\label{Labref}
\end{figure}

To compute the cross section we pick the lab frame as in Fig. \ref{Labref} \cite{Peskin}.
The photon has momentum $k ^a=(\omega, 0,0, \omega)$ and the electron is such that, in Lorentz indices, it reads 
\begin{equation}
\begin{pmatrix}
p^0 \\
p^1 \\
p^2 \\
p^3
\end{pmatrix} = \begin{pmatrix}
m+\frac{\alpha}{2r}\\
0\\
-m\alpha\sin{\theta}+\frac{1}{2r\sin{\theta}}\\
0 
\end{pmatrix}.
\end{equation}

To recover the free case, as in QFT, one should consider $\alpha \to 0$ and $r \to + \infty $. To the first order in $\alpha$, we have that $p^2=m^2 +2m \alpha / r-1/(2r \sin \theta)^2$. The term $2m \alpha / r$ is due to the potential energy of the hydrogen atom, while the term $1/(2r \sin \theta)^2$ is due to orbital angular momentum.

For the outgoing particles, the photon momentum is given by $ k^{' a} = ( \omega ', \omega ' \sin \theta ' , 0 , \omega ' \cos \theta ' )$ and the electron is free so that we can take $p'^2 = m^2 $ as usual. 

The cross section is
\begin{equation}
d \sigma = \frac{1}{2 \omega } \frac{1}{2(m + \alpha / r ) } d\Phi _2 | M(\{ k,p \} \rightarrow \{k', p' \} | ^2,
\end{equation}
as easily shown by textbook calculations \cite{Peskin}.

In order to calculate the explicit expression of $d \Phi _2 $ we will make use of the formula
\begin{eqnarray}
\int \frac{d^3 p'}{2 E_{p}}&=& \int d^3 p' dp' _0 \delta (p _0^{'2} - E_{p}^2) \Theta (p' _0 ) \nonumber \\
&=& \int d^3 p' dp'_0 \delta (p_0^{'2} - p^{'2}-\vec{p'}^2) \Theta (p'_0 ) \nonumber \\
&=& \int d^3 p' dp'_0 \delta (p'_a p^{'a} - p'^2) \Theta (p'_0 )
\end{eqnarray}
from which we get
\begin{align}
\nonumber
\int d \Phi _2= & \frac{1}{(2 \pi)^2} \int \frac{d^3 p'}{2 E_{p'}} \frac{d^3 k'}{2 \omega '} \delta ^{(4)} (k' + p' - k - p) \\
\nonumber
= & \frac{1}{(2 \pi)^2} \int \frac{d^3 k'}{2 \omega '} \delta ((p+k-k')^2 - m^2)\\
\nonumber
& \Theta (\omega + E_p - \omega ' )
=\frac{1}{2 (2 \pi)^2} \int \omega ' d \omega ' d \Omega\\
\nonumber
&\delta \bigg(2m \alpha / r-1/(2r \sin \theta)^2 + 2\omega (m + \alpha / r)\\
\nonumber
& - \omega ' [2(m + \alpha /r) + 2 (1- \cos \theta ' ) \omega\\ 
\nonumber
&+ \frac{\sin \varphi \sin \theta '}{r \sin \theta} ] \bigg)
=\frac{1}{4 \pi} \int d \cos \theta ' \\
& \times \frac{ \omega ^{'2}}{2m \alpha / r -1/(2r \sin \theta)^2 + 2\omega (m + \alpha / r) }
\end{align}
and eventually
\begin{eqnarray}
\frac{d \sigma}{d \cos \theta ' } = \frac{1}{2 \omega}\frac{1}{2( m + \alpha / r )} \mid M(\{ k,p \} \rightarrow \{k', p' \} ) \mid ^2 \nonumber \\
\times \frac{1}{4 \pi }\frac{\omega ^{'2}}{2m \alpha / r -1/(2r \sin \theta)^2 + 2\omega (m + \alpha / r)}, \quad \quad \quad \quad
\end{eqnarray}
where $|M|$ is the only unknown of the problem.

To evaluate $|M|$, the scalar products have to be explicitly performed. As we keep only the first order in $\alpha$ the total momentum $ q^a = p^a + k^a$ is
\begin{equation}
\begin{pmatrix}
q^0 \\
q^1 \\
q^2 \\
q^3
\end{pmatrix} = \begin{pmatrix}
m + \frac{\alpha}{2r} + \omega\\
0 \\
-m\alpha\sin{\theta}+\frac{1}{2r\sin{\theta}}\\
\omega
\end{pmatrix}.
\end{equation} 

Defining $ Y \equiv \cos \theta '$ in the limit $r \to \infty$ we obtain
\begin{align}
\frac{d \sigma}{ d Y}= \frac{\pi \alpha ^2}{m( m+ \omega - Y \omega)^3}
[ m^2 (1 + Y^2) \nonumber \\
+m(1-Y)(1+Y^2)\omega+(1-Y)^2 \omega^2].\label{cross}
\end{align}

Notice that when $\omega$ tends to zero, we find 
\begin{align}
\frac{d \sigma}{ d Y}= \frac{\pi \alpha ^2}{m^2}(1+Y^2)
\end{align}
which is the usual soft photon result of QFT. However, in general, the result is quite different, as a consequence of the interaction. Individually, if one photon hits an electron within the hydrogen potential, the probability of diffusion depends on where the electron is. In an hydrogen-like potential, $\beta$ is indeed non-vanishing and describes an internal dynamics between the left and the right parts of the electron, and this breaks the spherical symmetry. If, however, we assume a statistical phenomenon, the situation is different. For example, the spin-sum relationships used in (\ref{spinsum}) are statistical in nature and in this case, one needs to average the cross section (\ref{cross}) over $\theta$. The interaction term then cancels which, quite hopefully, leads to the usual result. Statistically, therefore, there is once again no correction term at the first order in $\alpha$ that seems to appear because of the electron-proton interaction.
%%%%%%%%%%%%%%%%%%%%%%%%%%%%%%%%%%%%%%%%%%%%%%%%%%%%%%%%%%%%%%%%%%%%%%%%%%%%%%%%%%%%%%%%%%%%%%%%%%%
%%%%%%%%%%%%%%%%%%%%%%%%%%%%%%%%%%%%%%%%%%%%%%%%%%%%%%%%%%%%%%%%%%%%%%%%%%%%%%%%%%%%%%%%%%%%%%%%%%%
\section{Conclusion}
In this paper, we have considered the polar form of spinors fields to obtain the most general propagator for the spinor fields themselves. Because such a propagator is the most comprehensive, then it should also contain all information about any possible interaction. Applying propagators in interaction means that we have already accounted for it and therefore one may wonder whether already at tree-level it is possible to recover the corrections normally obtained in QFT. To assess this issue, we have performed the computation of scattering amplitudes for the Bhabha scattering and the Compton scattering off a hydrogen-like potential. In the specific cases of these two instances, all corrections averaged out precisely as it would happen in standard QFT. What this means is that the question about whether or not interacting propagators at tree-level and free propagators with loop corrections can be equivalent seems to be not ruled out from these first examples. Of course more has to be done, and it would be best to find cases in which the equivalence seems to persist even when non-trivial corrections are found. We are developing such a case for the anomalous magnetic moment of the electron for a following paper.

However, there already are details that our general formalism seem to grasp: one of the most important is the fact that the small imaginary term $i\varepsilon$ that in standard QFT has to be introduced by hand to avoid poles, in the present formalism it elegantly appears on its own and it can be tied to the $x_{i}$ term suggesting that such a contribution accounts for the deviations of the matter distribution from being point-like. This is just another way of seeing that most of the singularity problems in QFT are due to the fact that the internal structure is neglected.

As it stands, it is notable the fact that dealing with such a general formalism can indeed bring advantages, both in alleviating problems or at least in rendering their solution less ad hoc as compared to standard QFT.

\

\noindent Competing Interests: NONE\\
Funding Info: NONE\\
Author contribution: all authors contributed\\
Consent to Participate: all authors gave their consent\\
Acknowledgements: NONE
%%%%%%%%%%%%%%%%%%%%%%%%%%%%%%%%%%%%%%%%%%%%%%%%%%%%%%%%%%%%%%%%%%%%%%%%%%%%%%%%%%%%%%%%%%%%%%%%%%%
%%%%%%%%%%%%%%%%%%%%%%%%%%%%%%%%%%%%%%%%%%%%%%%%%%%%%%%%%%%%%%%%%%%%%%%%%%%%%%%%%%%%%%%%%%%%%%%%%%%

%%%%%%%%%%%%%%%%%%%%%%%%%%%%%%%%%%%%%%%%%%%%%%%%%%%%%%%%%%%%%%%%%%%%%%%%%%%%%%%%%%%%%%%%%%%%%%%%%%%
\end{document}